\documentclass[final, number, 3p,12pt]{elsarticle}

\usepackage{amssymb}
\usepackage{amsthm}

\usepackage{mathrsfs}
\usepackage{float}
\usepackage{eufrak}
\usepackage{amsmath}
\usepackage{graphics}
\usepackage{amsfonts}
\usepackage{mathrsfs}
\usepackage{amscd}
\usepackage{color}
\usepackage{url}
\usepackage[plainpages=false,pdfpagelabels=true,colorlinks=true,citecolor=blue,hypertexnames=false]{hyperref}
\usepackage[vlined,ruled]{algorithm2e}

\usepackage{listings}
\journal{}

\newtheorem{theorem}{Theorem}[section]
\newtheorem{prop}[theorem]{Proposition}
\newtheorem{cor}[theorem]{Corollary}
\newtheorem{lemma}[theorem]{Lemma}

\newtheorem{define}[theorem]{Definition}
\newtheorem{example}[theorem]{Example}

\def\N{{\mathbb{N}}}
\def\Q{{\mathbb{Q}}}
\def\B{{\mathbb{B}}}
\def\G{{\mathbb{G}}}
\def\S{{\mathbb{S}}}
\def\H{{\mathbb{H}}}
\def\P{{\mathbb{P}}}
\def\M{{\mathbb M}}

\def\brm{{\bf \bar{m}}}

\def\z{{\bf 0}}

\def\lm{{\rm lm}} 
\def\lc{{\rm lc}}

\def\f{{\bf f}}
\def\t{{\bf t}}
\def\e{{\bf e}}

\def\u{{\bf u}}
\def\v{{\bf v}}
\def\w{{\bf w}}

\def\lcm{{\rm lcm}}

\def\deg{{\rm deg}}
\def\max{{\rm max}}
\def\min{{\rm min}}

\def\lla{{\longleftarrow}}

\newcommand{\comment}[1]{}
\newcommand{\ignore}[1]{}

\begin{document}

\begin{frontmatter}



\title{A Monomial-Oriented GVW for Computing Gr\"obner Bases
}



\author[sklois]{Yao Sun}
\ead{sunyao@iie.ac.cn}

\author[klmm]{Dingkang Wang}

\author[sklois]{Zhenyu Huang}

\author[sklois]{Dongdai Lin}

\address[sklois]{SKLOIS, Institute of Information Engineering, CAS, Beijing 100093,  China}

\address[klmm]{KLMM, Academy of Mathematics and Systems Science, CAS, Beijing 100190, China}

\begin{abstract}
The GVW algorithm, presented by Gao et al., is a signature-based algorithm for computing Gr\"obner bases. In this paper, a variant of GVW is presented. This new algorithm is called a monomial-oriented GVW algorithm or mo-GVW algorithm for short. The mo-GVW algorithm presents a new frame of GVW and regards {\em labeled monomials} instead of {\em labeled polynomials} as basic elements of the algorithm. Being different from the original GVW algorithm, for each labeled monomial, the mo-GVW makes efforts to find the smallest signature that can generate this monomial. The mo-GVW algorithm also avoids generating J-pairs, and uses efficient methods of searching reducers and checking criteria. Thus, the mo-GVW algorithm has a better performance during practical implementations. 
\end{abstract}


\end{frontmatter}



\section{Introduction} \label{sec_intro}

Gr\"obner bases, proposed by Buchberger in 1965 \citep{Buchberger65}, have been proven to be very useful in many aspects of algebra. In the past forty years, many efficient algorithms have been proposed to compute Gr\"obner bases. One important improvement is that Lazard pointed out the strong relation between Gr\"obner bases and linear algebra \citep{Lazard83}. This idea has been implemented  in F4 by Faug\`ere\citep{Fau99}, and also as XL type algorithms by Courtois et al. \citep{Courtois00} and Ding et al. \citep{Ding08}. 

Faug\`ere first introduced the concept of signatures for polynomials and presented the famous F5 algorithm \citep{Fau02}. Since then, signature-based algorithms have been widely investigated, and several variants of F5  have been presented, including F5C \citep{Eder10}, extended F5 \citep{Ars10}, F5 with revised criterion (the AP algorithm) \citep{Arri11}, and RB \citep{Roune13}.  Gao et al. proposed another signature based algorithm G2V \citep{Gao10a} in a different way from F5, and GVW\citep{Gao10b} is an extended version of G2V. The authors also studied generalized criteria and signature-based algorithms in solvable polynomial algebra in \citep{SunWang11b, SunWang12}.

In the field of implementations of signature-based algorithms, Faug\'ere presented his implementation of F5 in \citep{Fau02} and  improved it by parallel techniques in \citep{Fau10}.
A matrix-F5 is mentioned in \citep{Fau09,Bardet13}. 
An F5 algorithm in F4 style was described in more detail by Albrecht and Perry \citep{Albrecht10}. Roune and Stillman efficiently implemented  GVW and  AP without using linear algebra \citep{Roune12}. The authors implemented GVW in F4 style \citep{SunDW14} over boolean polynomial rings using routines modified from M4RI \citep{Albrecht13}.

When implementing GVW in F4 style \citep{SunDW14}, we find that, except the elimination of  matrices, some other procedures of GVW also cost very much time. The implementation of these procedures will affect the efficiency of the whole algorithm remarkably, particularly for complicated systems. These costly procedures include, generating J-pairs, searching reducers when doing top-reductions, and checking criteria, where reducers are the pairs/polynomials that are used to reduce others. To speed up the implementation of GVW, we present a new frame of the GVW algorithm in this paper, called a monomial-oriented GVW algorithm or mo-GVW algorithm for short.

We call this new algorithm a {\em monomial-oriented} algorithm, because labeled monomials instead of labeled polynomials are the basic elements of the mo-GVW algorithm. For each labeled monomial, mo-GVW makes effects to find the smallest signature that can generate this monomial. This is different from the original GVW algorithm, since GVW always tries to find the smallest monomial that can be generated by a given signature. 

In mo-GVW, J-pairs are not generated. Instead, monomials/polynomials are {\em lifted} similar like XL and matrix-F5. A criterion, named LCM Criterion, is used to avoid redundant computations during the lift. The mo-GVW algorithm also uses a new manner to find reducers and check criteria. In GVW, we often need to search a monomial from a large set such that this monomial divides some given monomial. This search may be very costly, since we have to traverse many monomials in a large set. The mo-GVW algorithm avoids this search, and instead, mo-GVW turns to check whether a monomial belongs to a large set. This new check can be easily done by using a hash table, and hence, saves much time. Using this method, mo-GVW can search reducers and check criteria very efficiently.

We implemented the mo-GVW algorithm over boolean polynomial rings. Efficient routines modified from M4RI \citep{Albrecht13} are mainly used to do one-side eliminations of matrices in mo-GVW, where the modification method is reported in \citep{SunDW14}. We tested our implementation of mo-GVW with many systems. The experimental results show mo-GVW is much more efficient than the M-GVW algorithm which is proposed in \citep{SunDW14}. Compared with some intrinsic Gr\"obner basis functions on public softwares, mo-GVW is also very efficient when the systems are not very complicated.

This paper is organized as follows. We introduce the theories of the mo-GVW algorithm in Section \ref{sec_theory}. We discuss the implementation of mo-GVW in Section \ref{sec_implementation}. Some experimental results are given in Section \ref{sec_timming}. Concluding remarks follow in Section \ref{sec_conclusion}.

\section{Theory} \label{sec_theory}

In this section, theories of the mo-GVW algorithm are presented. Necessary notations are given in Subsection \ref{subsec_notations}. A new data structure, called labeled monomials, is proposed in Subsection \ref{subsec_labeledmonomials}. We discuss reductions of mo-GVW in Subsection \ref{subsec_reduction}. A variant of strong Gr\"obner basis is defined in Subsection \ref{subsec_mosgb}. The mo-GVW algorithm comes in Subsection \ref{subsec_algorithm}. The relations between GVW and mo-GVW are discussed briefly in Subsection \ref{subsec_relation}. A toy example follows in Subsection \ref{subsec_example}.

\subsection{Notations} \label{subsec_notations}

Let $R:= k[X]$ be a polynomial ring over a field $k$ in variables $X = \{x_1, x_2, \ldots, x_n\}$. Given a finite vector of polynomials $\f = (f_1, f_2, \ldots, f_l) \subset R^l$, in this paper, we are going to compute a Gr\"obner basis for the ideal $$I:=\langle f_1, f_2, \ldots, f_l\rangle,$$ w.r.t. a monomial ordering $\prec_p$ on $R$.

Let $\M$ be the $R$-module $$\M:=\{(\u, f) \mid \u \cdot \f = p_1f_1+p_2f_2+\cdots+p_lf_l = f\in R, \u = (p_1, p_2, \ldots, p_l) \in R^l\}.$$ Clearly, $\M$ is generated by $\{(\e_1, f_1), (\e_2, f_2), \ldots, (\e_l, f_l)\}$ over $R$, where $\e_1, \e_2, \ldots, \e_l$ are the units in $R^l$. The notations of $I$ and $\M$ will be used throughout this paper.

Denote $Mon(R)$ and $Mon(R^l)$ be the set of all monomials in $R$ and $R^l$ respectively, i.e. $Mon(R) = \{x^\alpha = x_1^{a_1}x_2^{a_2}\cdots x_n^{a_n} \mid \alpha = (a_1, a_2, \ldots, a_n) \in \N^n\}$, where $\N$ is the set of all non-negative integers,  and $Mon(R^l) =\{x^\alpha\e_i \mid x^\alpha \in Mon(R)$ and $i = 1, 2, \ldots, l\}$.

Let $\prec_p$ be a monomial ordering on $R$, and $\prec_s$ be a {\bf position over term} extension of $\prec_p$ to $R^l$, i.e. $x^\alpha\e_i \prec_s x^\beta\e_j$, if either $i > j$, or $i = j$ and $x^\alpha \prec_p x^\beta$. For a polynomial $f\in R$ and a vector $\u\in R^l$, the leading monomial of $f$ and $\u$, denoted as $\lm(f)$ and $\lm(\u)$,  are defined as the largest monomials in $f$ and $\u$ w.r.t. the ordering $\prec_p$ and $\prec_s$ respectively. The leading coefficients of $f$ and $\u$, denoted as $\lc(f)$ and $\lc(\u)$, are the corresponding coefficients of $\lm(f)$ and $\lm(\u)$ in $f$ and $\u$ respectively. We make conventions that $\lm(0) = 0 \in Mon(R)$ and $\lm({\bf 0}) =  0 \in Mon(R^l)$. 
In this paper, we usually omit the subscripts of $\prec_p$ and $\prec_s$ if no confusions occur.

For any pair $(\u, f) \in \M$, we call $\lm(\u)$ the signature of $(\u, f)$. This definition of signature is the same as that in \citep{Gao10b}.

\subsection{Labeled monomials} \label{subsec_labeledmonomials}

Let $I$ and $\M$ be defined as the previous subsection. A monomial $m$ in $R$ is called an {\bf available leading monomial} w.r.t. $I$, if $m\in \lm(I) = \{\lm(f) \mid f \in I\}$.

\begin{define}
A vector $\brm = (m, (\u, f)) \in Mon(R) \times (\M\setminus \{(\z, 0)\})$ is called a {\bf labeled available leading monomial}  w.r.t. $I$ $(${\bf labeled monomial} for short$)$, if $\lm(f)$ divides $m$. 

Particularly, we say $\brm = (m, (\u, f))$ is {\bf primitive} if $m = \lm(f) \not= 0$, and $(0, (\u, 0))$ is called a {\bf syzygy} labeled monomial. 
\end{define}

Please note that $(0, (\z, 0))$ is not a labeled monomial.

For a labeled monomial $\brm = (m, (\u, f))$, the {\bf monomial}, {\bf generator}, {\bf degree}, and {\bf signature} of $\brm$ is defined as $m$, $(\u, f)$, $\deg(m)$, and $t\lm(\u)$ respectively, where $t = m/\lm(f)$ if $f\not= 0$, and $t = 1$ otherwise. Besides, we say $\brm$ is a labeled monomial of $m$. Please note that $t$ is usually not 1 in  mo-GVW.



We define the product of a monomial $x^\alpha$  and a labeled monomial $\brm = (m, (\u, f))$ as: $$x^\alpha \brm = (x^\alpha m, (\u, f)).$$ Clearly, $x^\alpha \brm$ is still a labeled monomial. In mo-GVW, we often need to {\bf lift a labeled monomial $\brm$ by $X = \{x_1, x_2, \ldots, x_n\}$}. That is, when we have obtained a labeled monomial $\brm$, we often need to consider the labeled monomials $x_1\brm$, $x_2\brm$, \ldots, $x_n\brm$ in the following steps.

In mo-GVW, it is possible that we obtained two labeled monomials $(m, (\u, f))$ and $(m', (\v, g))$ such that $m = m'\not=0$. We call this phenomenon as a {\bf collision} of labeled monomials, and say $(m, (\u, f))$ and $(m', (\v, g))$ {\bf collide} with each other. In this case, mo-GVW always retains only one labeled monomial of $m$. Specifically, the labeled monomial with a relative smaller {\em signature} is always retained, and the other one is discarded. That is, 

\begin{enumerate}

\item if $t\lm(\u) = t'\lm(\v)$, either one can be retained,

\item if $t\lm(\u) \succ t'\lm(\v)$, $(m', (\v, g))$ is retained,

\item if $t\lm(\u) \prec t'\lm(\v)$, $(m, (\u, f))$ is retained,

\end{enumerate}
where $t = m/\lm(f)$ and $t' = m'/\lm(g)$. In mo-GVW, we do not say sygyzy labeled monomials collide with others.



\begin{example}
Let $\f = (f_1, f_2) = (x + 1, y+2)$ be in $\Q[x, y]^2$ where $\Q$ is the rational field, and $\e_1 = (1, 0)$ and $ \e_2 = (0, 1)$. 

$(x, (\e_1, f_1))$ and $(y, (\e_2, f_2))$ are primitive labeled monomials w.r.t. $I = \langle f_1, f_2\rangle$. They can be lifted to $(x^2, (\e_1, f_1))$, $(xy, (\e_1, f_1))$, and $(xy, (\e_2, f_2))$, $(y^2, (\e_2, f_2))$, respectively. 

Note that the monomial $xy$ has two labeled monomials  $(xy, (\e_1, f_1))$ and $(xy, (\e_2, f_2))$, which is a collision. In mo-GVW, $(xy, (\e_2, f_2))$ is retained, since $(xy/\lm(f_2))\e_2 \prec (xy/\lm(f_1))\e_1$. 

Particularly, $(0,  (f_2\e_1 - f_1\e_2, 0))$ is a syzygy labeled monomial.

\end{example}

\subsection{Mutual-reductions} \label{subsec_reduction}

Let $\M$ and $I$ be defined as previous subsections. In this subsection, let $\brm = (m, (\u, f))$ be a labeled monomial, and $\B$ be a set of labeled monomials such that there are no collisions in $\B$, i.e. {\em any} two labeled monomials in $\B$ do not have the same monomial.

We say $\brm = (m, (\u, f))$ is {\bf reducible} by $\B$, if $\brm$ is {\em not} a syzygy labeled monomial and $\brm$ collides with $\brm' \in \B$ such that $\brm'$ has a {\em strictly smaller signature} than $\brm$, i.e. $f\not=0$, and there exists $\brm' = (m, (\v, g)) \in \B$ such that $t_f\lm(\u) \succ t_g\lm(\v)$ where $t_f = m/\lm(f)$ and $t_g =m/\lm(g)$. In this case, let $p = \lc(g)t_f f - \lc(f)t_g g$, and we say $\brm \longrightarrow (\lm(p), (\lc(g)t_f \u - \lc(f)t_g \v, p))$ is a one-step reduction. Please note that $t_f\lm(\u) = \lm(\lc(g)t_f \u - \lc(f)t_g \v)$, and $(\lm(p), (\lc(g)t_f \u - \lc(f)t_g \v, p))$ is either a primitive or a syzygy labeled monomial.

Assume $\brm$ is reducible by $\B$. We say $\brm$ is {\bf reduced} to $\brm''$ by $\B$, if $\brm''$ is not reducible by $\B$, and $\brm''$ is obtained by several one-step reductions from $\brm$ by $\B$. In this case, $\brm''$ has the following property, and its proof is directly from the definition.

\begin{prop} \label{prop_reduce}
If $\brm$ is reducible by $\B$ and $\brm$ is reduced to $\brm''$ by $\B$, then $\brm''$ is either a primitive or a syzygy labeled monomial. Besides, assume $\brm = (m, (\u, f))$ and $\brm''=(m'', (\w, h))$, we have $m \succ m''$ and $t_f\lm(\u) = \lm(\w)$ where $t_f = m/\lm(f)$.
\end{prop}

For convenience, we also say $\brm$ is reduced to $\brm$ by $\B$ if $\brm$ is not reducible by $\B$.

Please remark that, if $\brm$ is {\bf reduced} to $\brm''$ by $\B$, $\brm''$ may still collide with some $\brm' \in \B$. To deal with such collisions, mo-GVW does a {\em mutual-reduction} to $\brm$ by $\B$. We define the {\bf mutual-reduction} of $\brm$ by $\B$ in the following recursive way.

\begin{enumerate}[(i)]

\item Reduce $\brm$  to $\brm'' = (m'', (\w, h))$ by $\B$.

\item If $m''\not=0$ and $\brm''$ collides with $\brm' = (m'', (\v, g))$ in $\B$, then 

\begin{enumerate}[(A)]

\item If $t_h\lm(\w) \prec t_g\lm(\v)$ where $t_h = m''/\lm(h)$ and $t_g = m''/\lm(g)$, then 

\begin{enumerate}[(a)]

\item Let $\B \lla (\B \setminus \{\brm'\}) \cup \{\brm''\}$.

\item Mutual-reduce $\brm'$ by $\B$.

\end{enumerate}

\end{enumerate}

\item Otherwise, i.e. $m'' = 0$ or $\brm''$ does not collide with any $\brm' \in \B$,  let $\B \lla \B \cup \{\brm''\}$.

\end{enumerate}

Please note the following facts. The set $\B$ may be updated after mutual-reducing $\brm$.  In (i) we possibly have $\brm=\brm''$. In (ii), since $m''\not=0$ and $\brm''$ collides with $\brm' = (m'', (\v, g)) \in \B$, we must have $t_h\lm(\w) \preceq t_g\lm(\v)$, where $t_h = m''/\lm(h)$ and $t_g = m''/\lm(g)$. Particularly, if $t_h\lm(\w) = t_g\lm(\v)$ holds in (ii), nothing is done.

Mutual-reducing a labeled monomial can always be done within finite steps. Because the mutual-reduction is called recursively only when $\brm'$ is reducible by $\brm''$, and in this case, a labeled monomial with a strictly smaller monomial must appear in the following call of mutual-reduction. So the number of recursive calls is finite, since $\prec_p$ is a well-ordering.

In mo-GVW, by doing mutual-reduction to $\brm$ by $\B$, we aim to make labeled monomials in $\B$ have relative smaller signatures, and also ensure there are no collisions in $\B$.

\subsection{Monomial-oriented strong Gr\"obner bases} \label{subsec_mosgb}

Let $\M$ and $I$ be defined as previous subsections. In \citep{Gao10b}, a subset $G\subset \M$ is called a {\bf strong Gr\"obner basis} of $\M$, if any $(\u, f) \in \M$ is {\em top-reducible} by $G$, i.e. if $f=0$, there exists $(\v, 0)\in G$ such that $\lm(\v)$ divides $\lm(\u)$; otherwise, there exists $(\v, g)\in G$ such that $\lm(g)$ divides $\lm(f)$ and $\lm(\u) \succeq (\lm(f)/\lm(g)) \lm(\v)$. In mo-GVW, we need a definition of strong Gr\"obner bases for labeled monomials.

\begin{define}
Let $\G$ be a set of labeled monomials. The set $\G$ is called a {\bf monomial-oriented strong Gr\"obner basis} $(${\bf mo-strong Gr\"obner basis} for short$)$ of $\M$, if for any $0 = \u \cdot \f \in I$, there exists a syzygy labeled monomial $(0, (\v, 0)) \in \G$ such that $\lm(\v)$ divides $\lm(\u)$; and if for any $0\not= f = \u \cdot \f \in I$, there exists a primitive labeled monomial $(\lm(g), (\v, g)) \in \G$, such that 

\begin{enumerate}

\item $\lm(g)$ divides $\lm(f)$, and 

\item $\lm(\u) \succeq (\lm(f)/\lm(g))\lm(\v)$.

\end{enumerate}

\end{define}

The following proposition shows mo-strong Gr\"obner bases and strong Gr\"obner bases of $\M$ can be converted to each other easily. The proofs are trivial from the definitions.

\begin{prop} \label{prop_mosgb&sgb}
Let $\G$ be a set of labeled monomials and $G$ be a subset of $\M$.
\begin{enumerate}
\item If $\G$ is a mo-strong Gr\"obner basis of $\M$, then the set $\{(\u, f)\mid (\lm(f), (\u, f))\in \G\}$ is a strong Gr\"obner basis of $\M$.

\item If $G$ is a strong Gr\"obner basis of $\M$, then the set $\{(\lm(f),  (\u, f)) \mid (\u, f)\in G\}$ is a mo-strong Gr\"obner basis of $\M$.

\end{enumerate}
\end{prop}

\begin{cor}
If $\G$ is a mo-strong Gr\"obner basis of $\M$, then the set $\{f \mid (\lm(f), (\u, f)) \in \G\}$ is a Gr\"obner basis of $I$.
\end{cor}

\begin{proof}
Since the set $\{(\u, f) \mid (\lm(f), (\u, f))\in \G\}$ is a strong Gr\"obner basis of $\M$ by Proposition \ref{prop_mosgb&sgb}, the set $\{f \mid (\lm(f), (\u, f)) \in \G\}$ is a Gr\"obner basis by Proposition 2.2 in \citep{Gao10b}.
\end{proof}


Next, we modify Theorem 2.4 in \citep{Gao10b} slightly to present a labeled monomial version. 

First of all, we give the definitions of {\em J-pairs} and {\em cover}. Let  $(\u, f), (\v, g) \in \M$ with $fg\not=0$, a pair  $t_f(\u, f)$ is called the {\bf J-pair} of $(\u, f)$ and $(\v, g)$, if $t_f\lm(\u) \succ t_g\lm(\v)$ where $t_f = \lcm(\lm(f), \lm(g))/\lm(f)$ and $t_g = \lcm(\lm(f), \lm(g))/\lm(g)$. Particularly, if both $(\u, f)$ and $(\v, g)$ are in $G \subset \M$, we say $t_f(\u, f)$ is a {\em J-pair of $G$}. For a pair $(\u, f) \in \M$ and a set $G \subset \M$, we say $(\u, f)$ is {\bf covered} by $G$, if there is a pair $(\v, g)\in G$, such that $\lm(\v)$ divides $\lm(\u)$ and $t\lm(g) \prec \lm(f)$ $($strictly smaller$)$ where $t = \lm(\u)/\lm(\v)$.

\begin{lemma}[\textcircled{a} and \textcircled{c} of Thm. 2.4 in \citep{Gao10b}] \label{lem_gvw}
Suppose $G$ is a subset of $\M$ such that, for any monomial $\t \in Mon(R^l)$, there is a pair $(\v, g)\in G$ and a monomial $t$ such that $\t = t\lm(\v)$. Then $G$ is a strong Gr\"obner basis for $\M$ if and only if every J-pair of $G$ is covered by $G$.
\end{lemma}

In the following, we give definitions of {\em J-pairs} and {\em cover} in labeled monomial versions, and present a similar theorem afterwards.

Let $(\lm(f), (\u, f))$ and $(\lm(g), (\v, g))$ be two {\em primitive} labeled monomials. A labeled monomial $(m, (\u, f))$ is called the {\bf J-pair} of $(\lm(f), (\u, f))$ and $(\lm(g), (\v, g))$, if $m = \lcm(\lm(f), \lm(g))$ and $t_f\lm(\u) \succ t_g\lm(\v)$ where $t_f = m/\lm(f)$ and $t_g = m/\lm(g)$. Particularly, if both $(\lm(f), (\u, f))$ and $(\lm(g), (\v, g))$ are in a set $\G$, we say $(m, (\u, f))$ is a J-pair of $\G$. Please note that J-pairs of labeled monomials are only defined on {\em primitive} labeled monomials.

For a labeled monomial $(m, (\u, f))$ and a set $\G$ of labeled monomials, we say $(m, (\u, f))$ is {\bf covered} by $\G$, if there is a labeled monomial $(m', (\v, g))\in \G$, such that $\lm(\v)$ divides $t\lm(\u)$ and $t'\lm(g) \prec m$ $($strictly smaller$)$ where $t=m/\lm(f)$ and $t' = (t\lm(\u))/\lm(\v)$. Note that let $\B$ be a set of labeled monomials, if a labeled monomial $\brm$ is {\em reducible} by $\B$ and is {\em reduced} to $\brm''$ by $\B$, then $\brm$ is covered by $\{\brm''\}$.

The ``cover" relation is a one-side relation and has the transitivity, i.e. $\brm$ is covered by $\{\brm'\}$ never implies $\brm'$ is covered by $\{\brm\}$, and if $\brm$ is covered by $\{\brm'\}$ and $\brm'$ is covered by $\{\brm''\}$, then $\brm$ is covered by $\{\brm''\}$.

\begin{theorem}\label{thm_main}
Let $\G$ be a set of labeled monomials such that $(m, (\v, g)) \in \G$ implies $(\lm(g), (\v, g)) \in \G$, and for any monomial $\t \in Mon(R^l)$, there is $(m, (\v, g))\in \G$ and a monomial $t$ such that $\t = t\lm(\v)$. Then $\G$ is a mo-strong Gr\"obner basis for $\M$ if and only if every J-pair of $\G$ is covered by $\G$.
\end{theorem}

The proof of the above theorem is directly from Proposition \ref{prop_mosgb&sgb} and Lemma \ref{lem_gvw}. The above theorem deduces the following criteria.

\begin{cor}[{\bf Syzygy Criterion}] \label{cor_syzygy}
Let $(m, (\u, f))$ be a labeled monomial with $m\not=0$ and $\G$ be a set of labeled monomials. If there exists $(0, (\v, 0)) \in \G$ such that $\lm(\v)$ divides $t\lm(\u)$ where $t = m/\lm(f)$, then $(m, (\u, f))$ is covered by $\{(0, (\v, 0))\} \subset \G$, and hence, the labeled monomial $(m, (\u, f))$ does not need to be mutual-reduced.
\end{cor}

\begin{cor}[{\bf Rewritten Criterion}] \label{cor_rewritten}
Let $(m, (\u, f))$ be a labeled monomial and $\G$ be a subset of labeled monomials. If $(m, (\u, f))$ is covered by $\G$, then the labeled monomial $(m, (\u, f))$ does not need to be mutual-reduced.
\end{cor}

\begin{cor}[{\bf LCM Criterion}]\label{cor_lcm}
Let $\G$ be a set of labeled monomials,  $(\lm(f), (\u, f))$ be a primitive labeled monomial, and $t\in Mon(R)$. If there exists $(m = t\lm(f), (\v, g))\in \G$ such that 
\begin{enumerate}[(1)]
\item $t(\lm(f), (\u, f))$ is not the J-pair of $(\lm(f), (\u, f))$ and $(\lm(g), (\v, g))$, but is a multiple of this J-pair. That is, $t\lm(f)=m \not= \lcm(\lm(f), \lm(g))$ and $t\lm(\u) \succ t_g\lm(\v)$, where $t_g = m/\lm(g)$ 

\item the J-pair of $(\lm(f), (\u, f))$ and $(\lm(g), (\v, g))$ is covered by $\G$
\end{enumerate}
Then $t(\lm(f), (\u, f)) = (t\lm(f), (\u, f))$ is also covered by $\G$, and hence, the labeled monomial $(t\lm(f), (\u, f))$ does not need to be mutual-reduced.
\end{cor}

In mo-GVW, LCM Criterion is used when mutual-reducing a labeled monomial $(m, (\u, f))$ by a set $\G$ of labeled monomials. In this case, we say $(m, (\u, f))$ is {\bf rejected by LCM Criterion} if there exists $(m, (\v, g))\in \G$ such that condition (1) of Corollary \ref{cor_lcm} is met.

\subsection{The mo-GVW algorithm} \label{subsec_algorithm}

In this subsection, we present the mo-GVW algorithm. The following main ideas are used in mo-GVW.

\begin{enumerate}

\item A set $\G$ of labeled monomials is maintained in mo-GVW, such that there are no collisions  in $\G$, i.e. each nonzero monomial in $Mon(R)$ has at most one labeled monomial in $\G$.

\item A labeled monomial $\brm$ is inserted into $\G$, if (1) $\brm$ is a syzygy labeled monomial, or (2) $\brm$ does not collide with any labeled monomial in $\G$, or (3) $\brm$ collides with $\brm' \in \G$, but the signature of $\brm$ is smaller than the signature of $\brm'$.

\item A labeled monomial $\brm \in \G$ is {\bf lifted} to $x_1\brm$, $x_2\brm$, \ldots, $x_n\brm$, if $\brm$ is not a syzygy labeled monomial and $\brm$ has not been lifted yet.

\item LCM, Syzygy, and Rewritten Criterion are used in the mutual-reductions of labeled monomials to avoid redundant computations. 

\end{enumerate}



Next, we give the monomial-oriented GVW algorithm. To make proofs easier, we assume $\lm(f_i) \not= \lm(f_j)$ for $1 \le i < j \le l$.

\begin{algorithm}[H]
\DontPrintSemicolon
\SetAlgoSkip{}
\LinesNumbered

\SetKwData{maxdeg}{liftdeg}
\SetKwData{goto}{goto step}
\SetKwData{maxcpdeg}{maxcpdeg}
\SetKwFunction{mreduce}{mutualreduce}
\SetKwFunction{update}{update}
\SetKwFunction{return}{return}
\SetKwFunction{break}{break}
\SetKwInOut{Input}{Input}
\SetKwInOut{Output}{Output}
\SetKwFor{For}{for}{do}{end\ for}
\SetKwIF{If}{ElseIf}{Else}{if}{then}{else\ if}{else}{end\ if}

\Input{
$\{f_1, f_2, \ldots, f_l\}$, a finite subset of $R = k[x_1, x_2, \ldots, x_n]$, and $\lm(f_i) \not= \lm(f_j)$ for $1 \le i < j \le l$.


}

\Output{$\G$, A monomial-oriented strong Gr\"obner basis of $\M = \langle (\e_1, f_1), \ldots, (\e_l, f_l)\rangle$.}

\BlankLine

\Begin{

$\G \lla \{(0, (f_j\e_i - f_i\e_j,  0)) \mid 1 \le i < j \le l \} \cup \{(\lm(f_i), (\e_i, f_i)) \mid 1 \le i \le l\}$ 

$\maxdeg \lla \max\{\deg(\brm) \mid \brm \in \G$ is primitive$\}$

\While{$\exists\, \brm_0 \in \G$ s.t. $\brm_0$ is not a syzygy labeled monomial,  $\deg(\brm_0) \le \maxdeg$, and $\brm_0$ has not been lifted}
{

\For{$i \lla 1, 2, \ldots, n$}
{
$\mreduce(x_i\brm_0, \G)$
}

$\maxdeg \lla \max\{\deg(\brm) \mid \brm \in \G$ is primitive$\}$
}

$\maxcpdeg \lla \max\{\deg(\lcm(\lm(f), \lm(g))) \mid (\lm(f), (\u, f)), (\lm(g), (\v, g)) \in \G\}$

\If{$\maxcpdeg > \maxdeg + 1$}
{
$\maxdeg \lla \maxcpdeg - 1$

\goto 4
}

\return $\G$

}\caption{The Monomial-oriented GVW (mo-GVW) Algorithm}
\end{algorithm}

The degrees of labeled monomials in $\G$ are at most {\tt liftdeg} $+ 1$ throughout the algorithm. 

The procedure {\tt mutualreduce}$(\cdot)$ mutual-reduces a labeled monomial by a set of labeled monomials. Compared with the definition given in Subsection \ref{subsec_reduction}, the following procedure avoids some redundant computations by using criteria.

\begin{procedure}[H]
\DontPrintSemicolon
\SetAlgoSkip{}
\LinesNumbered

\SetKwFunction{mreduce}{mutualreduce}
\SetKwFunction{return}{return}
\SetKwInOut{Input}{Input}
\SetKwInOut{Output}{Object}
\SetKwFor{For}{for}{do}{end\ for}
\SetKwIF{If}{ElseIf}{Else}{if}{then}{else\ if}{else}{end\ if}

\Input{
$\brm = (m, (\u, f))$, a labeled monomial; 

\hspace{1.85cm}$\G$, a set of labeled monomials. 
}


\BlankLine

\Begin{

\If{$\brm$ is reducible by $\G$ and $\brm$ can be rejected by any of LCM, Syzygy and Rewritten Criterion}
{
\return
}

reduce $\brm$ to $\brm'' = (m'', (\w, h))$ by $\G$

\If{$m'' \not= 0$ and $\brm''$ collides with $\brm' = (m'', (\v, g)) \in \G$}
{

\If{$t_h\lm(\w) \prec t_g\lm(\v)$ where $t_h = m''/\lm(h)$, $t_g = m''/\lm(g)$}
{
$\G \lla (\G \setminus \{\brm'\}) \cup \{\brm''\}$

$\mreduce(\brm', \G)$
}
}
\Else
{
$\G \lla \G \cup \{\brm''\}$
}

}\caption{mutualreduce($\brm$, $\G$)}

\end{procedure}

We next prove the termination and correctness of the mo-GVW algorithm. The following lemma is needed in the proof of the termination.

\begin{lemma} \label{lem_primitive}
Only finite primitive labeled monomials can be generated in the mo-GVW algorithm, if $\prec_s$ and $\prec_p$ are compatible, i.e. $x^\alpha\e_i \prec_s x^\beta\e_i$ only if $x^\alpha \prec_p x^\beta$.
\end{lemma}


\begin{proof}
We define a map $\psi$ from primitive labeled monomials to $Mon(k[X, Y, Z])$ where $X=\{x_1, \ldots, x_n\}$, $Y = \{y_1, \ldots, y_n\}$, and $Z = \{z_1, \ldots, z_l\}$. That is, for any primitive labeled monomial $\brm =(m, (\u, f))$ with $m = \lm(f) = x^\alpha$ and $\lm(\u) = x^\gamma\e_i$, we define $\psi(\brm) = x^\alpha y^\gamma z_i$. Since $\brm$ is primitive, we have $m \not= 0$ and $\u \not= \z$, and hence, the map $\psi$ is well defined.

Let $\brm''_0 = (m'', (\w, h))$ be a primitive labeled monomial computed by mo-GVW, then $\brm''_0$  should be inserted into $\G$. Assume $\G_0$ is the set $\G$ just before $\brm''_0$ is inserted. We claim $\psi(\brm''_0)$ is {\em not} divisible by any $\psi(\brm')$, where $\brm'$ is a primitive labeled monomial in $\G_0$. This claim will prove the lemma, because the ideal generated by $\{\psi(\brm') \mid \brm'$ is primitive in $\G_0\}$ over $k[X, Y, Z]$ will be strictly enlarged after $\brm''_0$ is inserted into $\G_0$, and this ideal cannot be strictly enlarged infinitely.

We prove the claim by contradiction. Assume there exists a primitive labeled monomial $\brm'_0 = (m', (\v, g)) \in \G_0$ such that $\psi(\brm'_0)$ divides $\psi(\brm''_0)$. Then we have $\lm(g)$ divides $\lm(h)$ and $\lm(\v)$ divides $\lm(\w)$. Let $t = \lm(h)/\lm(g)$ and $s = \lm(\w) /\lm(\v)$.

First of all, we analyze how $\brm''_0$ is computed by mo-GVW. In mo-GVW, labeled polynomials can only be inserted to the set $\G$ at Line 7 and 10 of {\tt mutualreduce}$(\cdot)$. So by the procedure {\tt mutualreduce}$(\cdot)$, there exists a labeled monomial $\brm_0 = (m, (\u, f))$, such that $\brm''_0$ is reduced from $\brm_0$ by $\G_0$.

Next, we show $\brm_0$ is {\em reducible} by $\G_0$, and $\brm_0$ cannot be rejected by any of LCM, Syzygy and Rewritten Criterion. If $\brm_0$ is {\em not reducible} by $\G_0$, then we have $\brm''_0 = \brm_0$. Since $\brm''_0$ is primitive, $\brm_0$ is primitive. Note that $\brm_0$ and $\G_0$ are the input of {\tt mutualreduce}$(\cdot)$. However, the input $\brm$ of {\tt mutualreduce}$(\cdot)$ at Line 6 of mo-GVW is not primitive,  and the input $\brm$ of {\tt mutualreduce}$(\cdot)$ at Line 8 of {\tt mutualreduce}$(\cdot)$ is reducible by the input $\G$. This is a contradiction. So $\brm_0$ must be {\em reducible} by $\G_0$, and hence, $\brm_0$ cannot be rejected by any of LCM, Syzygy and Rewritten Criterion by the procedure {\tt mutualreduce}$(\cdot)$. 

At last, we show that both $t \not\prec s$ and $t \not \succeq s$, which is a contradiction. If $t \prec s$, then $\brm'_0 \in \G_0$ can be used to reduce $\brm''_0$ further, since $\lm(h) = t\lm(g)$ and $\lm(\w) = s\lm(\v) \succ t\lm(\v)$ by compatibility of $\prec_s$ and $\prec_p$. This contradicts with that fact $\brm''_0$ is reduced from $\brm_0$ by $\G_0$. So we have $t \not\prec s$. If $t \succeq s$, then $\brm_0$ should be rejected by Rewritten Criterion, since  $t_f\lm(\u) = \lm(\w) = s\lm(\v)$ and $m \succ m'' =  \lm(h) = t\lm(g) \succeq s\lm(g)$, where $t_f = m/\lm(f)$. Thus, we have $t \not\succeq s$. The claim is proved.
\end{proof}

\begin{theorem} \label{thm_termination}
The mo-GVW algorithm terminates in finite steps, if $\prec_s$ and $\prec_p$ are compatible, i.e. $x^\alpha\e_i \prec_s x^\beta\e_i$ only if $x^\alpha \prec_p x^\beta$.
\end{theorem}

\begin{proof}
In this proof, it suffices to show mo-GVW must terminate in finite steps if no new primitive labeled monomials are generated. This will prove the theorem, since by Lemma \ref{lem_primitive}, mo-GVW can only generated finite primitive labeled monomials.

Let $\G$ be the set of labeled monomials in the mo-GVW algorithm at some time, and $d$ be $\max\{\deg(\lcm(\lm(f)$, $\lm(g)))$ $\mid$ $(\lm(f), (\u, f)), (\lm(g), (\v, g)) \in \G\}$.  Assume no new primitive labeled monomials will be inserted into $\G$ from that time. Then the value of $\max\{d - 1, ${\tt liftdeg}$\}$ will not change. Note that the number of labeled monomials in the set $\{\brm_0\in \G \mid$ $\brm_0$ is not a syzygy labeled monomial, $\deg(\brm_0) \le$ $\max\{d - 1, ${\tt liftdeg}$\}$, and $\brm_0$ has not been lifted$\}$ is finite, and mutual-reducing a labeled monomial can be done in finite steps, which has been discussed in Subsection \ref{subsec_reduction}. To show mo-GVW terminates in finite steps, it suffices to show only finite $\brm''$s will be inserted into $\G$ in the following steps, where $\brm''$ is not a syzygy labeled monomial and $\deg(\brm'') \le$ $\max\{d - 1, ${\tt liftdeg}$\}$.

By mo-GVW, a labeled monomial $\brm'' = (m'', (\w, h))$ can be inserted into $\G$ only at Line 7 and 10 of {\tt mutualreduce}$(\cdot)$. If $\brm''$ is inserted into $\G$ at Line 7, then the signature of $m''$'s labeled monomial in $\G$ will be strictly lowered; if $\brm''$ is inserted into $\G$ only at Line 10, then either $\brm''$ is a syzygy labeled monomial or the number of non-syzygy labeled monomials in $\G$ strictly increases. Because the ordering $\prec_s$ on signatures is a well-ordering and the number of nonzero monomials with degree not bigger than $\max\{d - 1, ${\tt liftdeg}$\}$ is finite. Only finite non-syzygy labeled monomial $\brm''$'s with $\deg(\brm'') \le$ $\max\{d - 1, ${\tt liftdeg}$\}$, will be inserted into $\G$. Thus, mo-GVW must terminate in finite steps.
\end{proof}

Next, we prove the correctness of the mo-GVW algorithm. The following lemmas and corollary are needed.

\begin{lemma} \label{lem_cover}
In the mo-GVW algorithm, let $\brm'$ be a primitive labeled monomial in the set $\G$. If $\brm'$ is removed from $\G$ at some time and $\G$ is updated to $\G'$, then $\brm'$ is covered by $\G'$. 
\end{lemma}

\begin{proof}
In mo-GVW, a labeled monomial $\brm' = (m', (\v, g))$ can only be removed from $\G$ at Line 7 of the procedure {\tt mutualreduce}$(\cdot)$. At this time, there must be a labeled monomial $\brm'' = (m'', (\w, h))$ such that $\brm''$ and $\brm'$ have the same monomial, but $\brm''$ has a strictly smaller signature than $\brm'$. In the following Line 8, the next step is to mutual-reduce $\brm'$ by $\G_0 = (\G \cup \{\brm''\}) \setminus \{\brm'\}$.

During the mutual-reduction of $\brm'$ by $\G_0$, $\brm'$ is reducible by $\{\brm''\} \subset \G_0$. Note that $\brm'$ cannot be rejected by LCM Criterion since $\brm'$ is primitive. If $\brm'$ is rejected by Syzygy or Rewritten Criterion, then by Corollary \ref{cor_syzygy} or \ref{cor_rewritten}, $\brm'$ is covered by $\G_0$, and hence, the mutual-reduction of $\brm'$ by $\G_0$ is over, which means we have $\G' = \G_0$ and the lemma is proved. Otherwise, $\brm'$ is reduced to a labeled monomial $\brm''_0$ by $\G_0$, and $\brm''_0$ is either a primitive or a syzygy labeled monomial by Proposition \ref{prop_reduce}. In either case, $\brm'$ is covered by $\{\brm''_0\}$, since $\brm'$ is reducible by $\G_0$. 

If either $\brm''_0$ is a syzygy labeled monomial,  or $\brm''_0$ is primitive but does not collide with any labeled monomial in $\G_0$, then $\brm''_0$ will be inserted into $\G_0$, and the mutual-reduction is over with $\G' = \G_0 \cup \{\brm''_0\}$. Then, $\brm'$ is covered by $\{\brm''_0\} \subset \G'$ and the lemma is proved.

If $\brm''_0$ is primitive and collides with $\brm'_0 \in \G_0$, then two possible cases happen. (a) if $\brm''_0$ and $\brm'_0$ have the same signature, then nothing is done and the mutual-reduction is over with $\G' = \G_0$. Since $\brm''_0$ and $\brm'_0$ have the same monomial and signature, $\brm'$ is covered by $\{\brm'_0\} \subset \G'$. (b) if the signature of $\brm''_0$ is strictly smaller than that of $\brm'_0$, then $\brm''_0$ is inserted into $\G_0$, and hence, $\brm''_0\in \G'$ because $\brm''_0$ cannot be removed from $\G'$ during the mutual-reduction of $\brm'_0$ by $(\G_0 \setminus \{\brm'_0\}) \cup \{\brm''_0\}$. So $\brm'$ is covered by $\{\brm''_0\} \subset \G'$. 
\end{proof}

\begin{cor} \label{cor_trans}
Let $\G_{end}$ be the output of the mo-GVW algorithm. If a labeled monomial $\brm$ is covered by some $\G$ during the computation of mo-GVW, then $\brm$ is covered by $\G_{end}$.
\end{cor}

\begin{proof}
Assume $\brm$ is covered by $(m, (\v, g)) \in \G$. If $(m, (\v, g)) \in \G_{end}$ or $(\lm(g), (\v, g)) \in \G_{end}$, then the lemma is proved. Otherwise, $(\lm(g), (\v, g))$ is removed from a set $\G_0$ at some time, and $\G_0$ is updated to $\G'_0$. By Lemma \ref{lem_cover}, $(\lm(g), (\v, g))$ is covered by $\G'_0$. Next, we can discuss whether $(\lm(g), (\v, g))$ is covered by $\G_{end}$ in a similar way. That is, assume $(\lm(g), (\v, g))$ is covered by $(m', (\w, h)) \in \G'_0$. If $(m', (\w, h)) \in \G_{end}$ or $(\lm(h), (\w, h)) \in \G_{end}$, then $(\lm(g), (\v, g))$ is covered by $\G_{end}$ and hence, $\brm$ is covered by $\G_{end}$. Otherwise, $(\lm(h), (\w, h))$ is removed from a set $\G_1$ at some time, and $\G_1$ is updated to $\G'_1$. By Lemma \ref{lem_cover}, $(\lm(h), (\w, h))$ is covered by $\G'_1$. In the following steps,  we can discuss whether $(\lm(h), (\w, h))$ is covered by $\G_{end}$ repeatedly. If the above discussions are infinite, then we can construct an array of primitive labeled monomials: $(\lm(g), (\v, g))$, $(\lm(h), (\w, h))$, ....  Each primitive labeled monomial in this array is covered by the successive one, and none of these primitive labeled monomials lies in $\G_{end}$.

Note that the ``cover" relation is a one-side relation, and mo-GVW terminates in finite steps, so only finite primitive labeled monomials have been removed from the set $\G$ during the computation. Therefore, the discussions in the last  paragraph cannot be infinite. We must have $\brm$ is covered by $\G_{end}$. 
\end{proof}

\begin{lemma} \label{lem_implie}
Let $\G_{end}$ be the output of the mo-GVW algorithm. 
\begin{enumerate}[(1)]

\item If $(m, (\v, g))$ is a labeled monomial obtained by mo-GVW during the computation,  and $t$ is a monomial such that $tm$ has a labeled monomial $\brm$ in $\G_{end}$, then the signature of $\brm$ is not bigger than the signature of $(tm, (\v, g))$.

\item $(m, (\v, g))\in \G_{end}$ implies $(\lm(g), (\v, g))\in \G_{end}$. 

\end{enumerate}
\end{lemma}

\begin{proof}

For (1), since $m$ divides $tm$, $(m, (\v, g))$ can be lifted to $(tm, (\v, g))$. If $(tm, (\v, g)) \in \G_{end}$, then (1) is proved. Otherwise, the lift from $(m, (\v, g))$ to $(tm, (\v, g))$ must be interrupted. That is, there exists $t'$ such that $t'$ divides $t$, $t' \not= t$, and $t'(m, (\v, g))$ is reducible by $\G$ at some time. In this case, there exists $(m', (\w, h)) \in \G$ such that $m' = t'm$ and the signature of $(m', (\w, h))$ is strictly smaller than $t'(m, (\v, g))$. After mutual-reducing $t'(m, (\v, g))$, the labeled monomial $(m', (\w, h))$ will be lifted in the following computations   instead of $t'(m, (\v, g))$. Note that $m' = t'm$ divides $tm$, $(m', (\w, h))$ can also be lifted to $(tm, (\w, h))$. Next, we can discuss whether $(tm, (\w, h))$ is in $\G_{end}$ similarly. This discussion can be repeated until we find $tm$'s labeled monomial in $\G_{end}$. The discussions are not infinite, since $\prec_s$ is a well-ordering and the signatures of $(tm, (\v, g))$, $(tm, (\w, h))$, ..... decrease strictly. Finally, (1) is proved.

For (2), if $(m, (\v, g))\in \G_{end}$ and $(\lm(g), (\v, g)) \not\in \G_{end}$, then $(\lm(g), (\v, g))$ must be removed from the set $\G$ at some time during the computation. Since $(\lm(g), (\v, g))$ can only be removed from $\G$ at Line 7 of {\tt mutualreduce}$(\cdot)$, $\lm(g)$'s labeled monomial in $\G_{end}$ must have a strictly smaller signature than $(\lm(g), (\v, g))$, i.e. there exists $(m'', (\w, h)) \in \G_{end}$ such that $m'' = \lm(g)$ and $t_h\lm(\w) \prec \lm(\v)$ where $t_h = m''/\lm(h)$. Since $m'' = \lm(g)$ divides $m$, $(m'', (\w, h))$ can be lifted to $(m, (\w, h))$. So the signature of $m$'s labeled monomial in $\G_{end}$ is not bigger than the signature of $(m, (\w, h))$ by (1), and hence, is strictly smaller than the signature of $(m, (\v, g))$. This is a contradiction with $(m, (\v, g))\in \G_{end}$. So $(m, (\v, g))\in \G_{end}$ implies $(\lm(g), (\v, g))\in \G_{end}$.
\end{proof}

\begin{theorem} \label{thm_correctness}
The mo-GVW algorithm computes a monomial-oriented strong Gr\"obner basis.
\end{theorem}

\begin{proof}
Let $\G_{end}$ be the output of the mo-GVW algorithm. To show $\G_{end}$ is a monomial-oriented strong Gr\"obner basis by Theorem \ref{thm_main}, it suffices to show that (1) $(m, (\v, g))\in \G_{end}$ implies $(\lm(g), (\v, g))\in \G_{end}$, which has been proved by Lemma \ref{lem_implie}, (2) for any monomial $\t \in Mon(R^l)$, there is $(m, (\v, g)) \in \G_{end}$ and a monomial $t$ such that $\t = t\lm(\v)$, and (3) every J-pair of $\G_{end}$ is covered by $\G_{end}$. 

For (2), the labeled monomial $(\lm(f_i), (\e_i, f_i))$ is inserted into $\G$ at Line 2 of mo-GVW. If $(\lm(f_i), (\e_i, f_i)) \in \G_{end}$ for all $1 \le i \le l$, then (2) is proved. Otherwise, some $(\lm(f_i), (\e_i, f_i))$ must be removed from the set $\G$ at some time during the computation. Since $(\lm(f_i), (\e_i, f_i))$ is primitive, by Lemma \ref{lem_cover} and  Corollary \ref{cor_trans}, $(\lm(f_i), (\e_i, f_i))$ is covered by $\G_{end}$. Then there exists $(m, (\u, f)) \in \G_{end}$ such that $\lm(\u)$ divides $\e_i$. In this case, $\lm(\u)$ has to be $\e_i$. Then (2) is proved.

For (3), let {\tt JPair} be the set of all J-pairs generated by primitive labeled monomials in $\G_{end}$. Clearly, {\tt JPair} is finite. We show all J-pairs in {\tt JPair} are covered by $\G_{end}$ by induction on an order of J-pairs, i.e. we say a J-pair $t_1(\lm(f), (\u, f))$ is {\em smaller} than a J-pair $t_2(\lm(g), (\v, g))$, if either $t_1\lm(\u) \prec t_2\lm(\v)$, or $t_1\lm(\u) = t_2\lm(\v)$ and $t_1\lm(f) < t_2\lm(g)$. 

Let $t(\lm(f), (\u, f))$ be a J-pair in {\tt JPair} where $(\lm(f), (\u, f))$ is a primitive labeled monomial in $\G_{end}$. Assume all J-pairs in {\tt JPair} that are (strictly) smaller than $t(\lm(f), (\u, f))$ have been shown covered by $\G_{end}$, we next show $t(\lm(f), (\u, f))$ is also covered by $\G_{end}$. Due to Line 8 to 11 of mo-GVW, we have $\deg(t\lm(f)) \le $ {\tt liftdeg} $+1$ when mo-GVW terminates. So only two possible cases can happen: (a) $t(\lm(f), (\u, f))$ is practically mutual-reduced in mo-GVW, and (b) $t(\lm(f), (\u, f))$ is not mutual-reduced in mo-GVW.

For case (a), when doing mutual-reduction\footnote{It is possible that $t(\lm(f), (\u, f))$ is mutual-reduced several times during the computation. Here we mean the last time of mutual-reduction.}, $t(\lm(f), (\u, f))$ is reducible by the set $\G$ at that time by the definition of J-pairs. If $t(\lm(f), (\u, f))$ is rejected by LCM, then $t(\lm(f), (\u, f))$ is covered by $\G_{end}$ by the inductive assumption and Corollary \ref{cor_lcm}. If $t(\lm(f), (\u, f))$ is rejected by Syzygy or Rewritten Criterion, then according to Corollary \ref{cor_syzygy} and \ref{cor_rewritten}, $t(\lm(f), (\u, f))$ is covered by $\G$,  and hence, it is covered by $\G_{end}$ by Corollary \ref{cor_trans}. Otherwise, $t(\lm(f), (\u, f))$ is reduced to $\brm''$, and $t(\lm(f), (\u, f))$ is covered by $\{\brm''\}$. With a similar discussion as the last two paragraphs in the proof of Lemma \ref{lem_cover}, we can show $t(\lm(f), (\u, f))$ is covered by $\G'$, where $\G'$ is the set after mutual-reducing $t(\lm(f), (\u, f))$ by $\G$. Then by Corollary \ref{cor_trans}, we have $t(\lm(f), (\u, f))$ is covered by $\G_{end}$.

Generally, since $(\lm(f), (\u, f)) \in \G_{end}$ and $\lm(f)$ divides $t\lm(f)$, the labeled monomial should be lifted to $(t\lm(f), (\u, f))$ and then $(t\lm(f), (\u, f))$ is mutual-reduced. However, in mo-GVW, the lift from $(\lm(f), (\u, f))$ to $(t\lm(f), (\u, f))$ may be interrupted sometimes. That is, there may exist $t'$ such that $t'$ divides $t$, $t' \not= t$, and $t'(\lm(f), (\u, f))$ is reducible by $\G$ at some time. In this case, there exists $(m, (\v, g)) \in \G$ such that $m = t'\lm(f)$ and the signature of $(m, (\v, g))$ is strictly smaller than $t'(\lm(f), (\u, f))$. After mutual-reducing $t'(\lm(f), (\u, f))$, the labeled monomial $(m, (\v, g))$ will be lifted in the following computations instead of $t'(\lm(f), (\u, f))$. So case (b) probably happens in mo-GVW.

For case (b), there must exist $t'$ such that $t'$ divides $t$, $t' \not= t$, $t'(\lm(f), (\u, f))$ is reducible by the set $\G$ as some time during the algorithm. So the monomial $t'\lm(f)$ must have a labeled monomial $(t'\lm(f), (\w, h)) \in \G_{end}$ and the signature of $(t'\lm(f), (\w, h))$ is smaller than the signature of $t'(\lm(f), (\u, f))$ by Lemma \ref{lem_implie}. Besides, we have $(\lm(h), (\w, h))\in \G_{end}$. Then $t'(\lm(f), (\u, f))$ is a multiple of the J-pair of $(\lm(f), (\u, f))$ and $(\lm(h), (\w, h))$. Since the signature of the J-pair of $(\lm(f), (\u, f))$ and $(\lm(h), (\w, h))$ is not bigger than the signature of $t'(\lm(f), (\u, f))$, and the signature of $t'(\lm(f), (\u, f))$ is strictly smaller than the signature of $t(\lm(f), (\u, f))$, the J-pair of $(\lm(f), (\u, f))$ and $(\lm(h), (\w, h))$ is smaller than $t(\lm(f), (\u, f))$, and hence, is covered by $\G_{end}$ due to the inductive assumption. So $t'(\lm(f), (\u, f))$ and hence $t(\lm(f), (\u, f))$ is covered by $\G_{end}$.
\end{proof}

\subsection{GVW and mo-GVW} \label{subsec_relation}

GVW can be regarded as a signature-oriented algorithm, while mo-GVW is a monomial-oriented algorithm. Reductions in GVW aims to find the smallest leading monomials for given signatures. By doing mutual-reductions, mo-GVW aims to find the smallest signatures for given monomials.

Criteria used in GVW and mo-GVW are the same. In GVW, LCM Criterion is used when generating J-pairs.

The number of non-syzygy pairs in the output of GVW is generally larger than the number of primitive labeled monomials in the output of mo-GVW. This is because many primitive labeled monomials are removed from the set $\G$ during the computation of mo-GVW.

\subsection{A toy example} \label{subsec_example}

\begin{example}
Let $F=\{f_1, f_2, f_3\} \subset \mathbb{F}_5[a, b, c]$, where $\mathbb{F}_5$ is the finite field $GF(5)$, $f_1 = abc-1$, $f_2=ab-c$, and $f_3=bc-b$. $\prec_p$ is the Graded Reverse Lex ordering with $a > b > c$, and $\prec_s$ is a position over term extension of $\prec_p$ with $\e_1 \succ_s \e_2 \succ_s \e_3$.
\end{example}

We compute a mo-strong Gr\"obner basis for $\M=\langle r_1 = (\e_1, f_1), r_2 = (\e_2, f_2), r_3 = (\e_3, f_3) \rangle$ by the mo-GVW algorithm.

Initially, {\tt liftdeg} $ = 3$ and $\G = \{(abc, r_1), (ab, r_2), (bc, r_3)\} \cup \S$, where $\S = \{(0, (ab \e_1 - abc \e_2, 0))$, $(0, (bc\e_1 - abc\e_3, 0))$, $(0, (bc\e_2 - ab\e_3, 0))\}$. 

\smallskip {\bf LOOP 1:} We choose $(bc, r_3)$ to lift. 

Multiplied by $c$ and $b$, $(bc, r_3)$ is lifted to $(bc^2, r_3)$ and $(b^2c, r_3)$, which are inserted into $\G$ directly. 

By multiplying $a$, we get $(abc, r_3)$. There exists $(abc, r_1) \in \G$ and the signature of $(abc, r_3)$ is smaller, i.e. $a\e_3 \prec \e_1$. By the procedure {\tt mutualreduce($\cdot$)}, $(abc, r_3)$ replaces $(abc, r_1)$ as the labeled monomial of $abc$ in $\G$. Then, after mutual-reducing $(abc, r_1)$, we achieve a new labeled monomial $(c, r_4)$, where $r_4 = (\e_1-\e_2-a\e_3, c-1)$. Then $(c, r_4)$ is inserted into $\G$.

After LOOP 1, we have $\G = \{(abc, r_3), (b^2c, r_3), (bc^2, r_3), (ab, r_2), (bc, r_3)$, $(c, r_4)\}$ $\cup$ $\S$ and {\tt liftdeg} $=2$.

\smallskip {\bf LOOP 2:} We choose $(c, r_4)$ to lift. 

Multiplied by $c$, $(c, r_4)$ is lifted to $(c^2, r_4)$, which is inserted into $\G$ directly. 

By multiplying $b$, we get $(bc, r_4)$. There exists $(bc, r_3) \in \G$ and the signature of $(bc, r_3)$ is smaller. According to the procedure {\tt mutualreduce($\cdot$)}, the labeled monomial $(0, (b\e_1 - b\e_2 -ab\e_3 - \e_3, 0))$ is generated and inserted into $\S \subset \G$.

Multiplied by $a$, $(c, r_4)$ is lifted to $(ac, r_4)$, which is inserted into $\G$ directly. 

After LOOP 2, we have $\G = \{(abc, r_3), (b^2c, r_3), (bc^2, r_3), (ab, r_2), (ac, r_4)$, $(bc, r_3), (c^2, r_4)$, $(c, r_4)\}$ $\cup$ $\S$ and {\tt liftdeg} $=2$, where $\S = \{(0, (ab\e_1 - abc\e_2, 0))$, $(0, (bc\e_1 - abc\e_3, 0))$, $(0, (bc\e_2 - ab\e_3, 0)), (0, (b\e_1 - b\e_2 -ab\e_3 - \e_3, 0))\}$.

\smallskip {\bf LOOP 3:} We choose $(ac, r_4)$ to lift. 

Multiplied by $c$, $(ac, r_4)$ is lifted to $(ac^2, r_4)$, which is inserted into $\G$ directly. 

By multiplying $b$, we get $(abc, r_4)$. There exists $(abc, r_3) \in \G$ and the signature of $(bc, r_3)$ is smaller. However, $(abc, r_4)$ is not a J-pair, and hence it is rejected by LCM Criterion and no reduction is done.

Multiplied by $a$, $(ac, r_4)$ is lifted to $(a^2c, r_4)$, which is inserted into $\G$ directly. 

After LOOP 3, we have $\G = \{(a^2c, r_4)$, $(abc, r_3)$, $(b^2c, r_3)$, $(ac^2, r_4)$, $(bc^2, r_3)$, $(ab, r_2)$, $(ac, r_4)$, $(bc, r_3)$, $(c^2, r_4)$, $(c, r_4)\}$ $\cup$ $\S$ and {\tt liftdeg} $=2$.

\smallskip {\bf LOOP 4:} We choose $(ab, r_2)$ to lift. 

By multiplying $c$, we get $(abc, r_2)$. There exists $(abc, r_3) \in \G$ and the signature of $(abc, r_3)$ is smaller. After reducing $(abc, r_2)$ by $\G$, we obtain $(c^2, r_5)$ where $r_5 = (c\e_2-\e_2-a\e_3, -c^2+c)$. There exists $(c^2, r_4) \in \G$ but the signature of $(c^2, r_5)$ is smaller. So $(c^2, r_5)$ is inserted into $\G$ and $(c^2, r_4)$ is removed. Next, we obtain the labeled monomial $(0, (c\e_1 - \e_2 - ac\e_3 - a\e_3, 0))$ by doing mutual-reduction to $(c^2, r_4)$. This labeled monomial is inserted into $\S \subset \G$.

Multiplied by $b$ and $a$, $(ab, r_2)$ is lifted to $(ab^2, r_2)$ and $(a^2b, r_2)$, which are inserted into $\G$ directly. 

After LOOP 4, we have $\G = \{(a^2b, r_2)$, $(ab^2, r_2)$, $(a^2c, r_4)$, $(abc, r_3)$, $(b^2c, r_3)$, $(ac^2, r_4)$, $(bc^2, r_3)$, $(ab, r_2)$, $(ac, r_4)$, $(bc, r_3)$, $(c^2, r_5)$, $(c, r_4)\}$ $\cup$ $\S$ and {\tt liftdeg} $=2$, where $\S = \{(0, (ab\e_1 - abc\e_2, 0))$, $(0, (bc\e_1 - abc\e_3, 0))$, $(0, (bc\e_2 - ab\e_3, 0))$, $(0, (b\e_1 - b\e_2 -ab\e_3 - \e_3, 0))$, $(0, (c\e_1 - \e_2 - ac\e_3 - a\e_3, 0))\}$.

\smallskip {\bf LOOP 5:} We choose $(c^2, r_5)$ to lift. 

Multiplied by $c$, $(c^2, r_5)$ is lifted to $(c^3, r_5)$, which is inserted into $\G$ directly. 

By multiplying $b$, we get $(bc^2, r_5)$. There exists $(bc^2, r_3) \in \G$ and the signature of $(bc^2, r_3)$ is smaller. Note that the signature of $(bc^2, r_5)$ is $bc\e_2$. No reduction is done due to Syzygy Criterion, since $(0, (bc\e_2 - ab\e_3, 0))\in \S$.

By multiplying $a$, we get $(ac^2, r_5)$. There exists $(ac^2, r_4) \in \G$ and the signature of $(ac^2, r_5)$ is smaller. So $(ac^2, r_4)$ is removed from $\G$ and $(ac^2, r_5)$ replaces it. Next, since $(ac^2, r_4)$ is not a J-pair, no reduction is done.

After LOOP 5, we have $\G = \{(a^2b, r_2)$, $(ab^2, r_2)$, $(a^2c, r_4)$, $(abc, r_3)$, $(b^2c, r_3)$, $(ac^2, r_5)$, $(bc^2, r_3)$, $(c^3, r_5)$, $(ab, r_2)$, $(ac, r_4)$, $(bc, r_3)$, $(c^2, r_5)$, $(c, r_4)\}$ $\cup$ $\S$ and {\tt liftdeg} $=2$.

So far, there is no unlifted labeled monomials in $\G$ with degrees not bigger than {\tt liftdeg} $=2$. Besides, for current $\G$, the value of {\tt maxcpdeg} is 3, and the mo-GVW algorithm is over. The set of primitive and syzygy labeled monomials in $\G$ is 
$$\{(ab, r_2), (bc, r_3), (c^2, r_5), (c, r_4)\} \cup \S.$$ This set is a mo-strong Gr\"obner basis of $\M$, and hence, $\{ab-c, bc-b, c-1, -c^2+c\}$ is a Gr\"obner basis of $\langle f_1, f_2, f_3\rangle$.

\section{On implementing the mo-GVW algorithm} \label{sec_implementation}

In this subsection, we talk about some implementing details of the mo-GVW algorithm. First, we rewrite the mo-GVW algorithm in matrix-style in Subsection \ref{subsec_matrixmogvw}; second, we show how to check Syzygy and Rewritten Criterion efficiently in Subsection \ref{subsec_checkcriterion}; some other details are discussed in Subsection \ref{subsec_others}.

\subsection{The matrix mo-GVW algorithm} \label{subsec_matrixmogvw}

To use the efficient techniques from linear algebra, it is necessary to put lots of reductions together and then do all these reductions at the same time. We revise the mo-GVW algorithm in matrix-style, and get the following algorithm. In the following algorithm, a labeled monomial $(m, (\u, f))$ is simply stored as $(m, (\lm(\u), f))$ instead of the whole $(m, (\u, f))$, since the information of $\u - \lc(\u)\lm(\u)$ is not useful throughout the algorithm. Similar operations is also done in \citep{Gao10b}.

\begin{algorithm}[H]
\DontPrintSemicolon
\SetAlgoSkip{}
\LinesNumbered

\SetKwData{todo}{todo}
\SetKwData{goto}{goto step}
\SetKwData{mindeg}{mindeg}
\SetKwData{liftdeg}{liftdeg}
\SetKwFunction{lift}{lift}
\SetKwFunction{mreduce}{mutual-reduce}
\SetKwFunction{append}{append}
\SetKwFunction{eliminate}{eliminate}
\SetKwFunction{update}{update}
\SetKwFunction{maxcpdeg}{maxcpdeg}
\SetKwFunction{return}{return}
\SetKwInOut{Input}{Input}
\SetKwInOut{Output}{Output}
\SetKwFor{For}{for}{do}{end\ for}
\SetKwIF{If}{ElseIf}{Else}{if}{then}{else\ if}{else}{end\ if}

\Input{
$\{f_1, f_2, \ldots, f_l\}$, a finite subset of $R = k[x_1, x_2, \ldots, x_n]$, and $\lm(f_i) \not= \lm(f_j)$ for $1 \le i < j \le l$.


}

\Output{A Gr\"obner basis of $\langle f_1, \ldots, f_m\rangle$.}
\BlankLine

\Begin{

$\G \lla \{(\lm(f_i), (\e_i, f_i)) \mid 1 \le i \le l\}$


%

$\liftdeg \lla \max\{\deg(\brm) \mid \brm \in \G$ is primitive$\}$

$\mindeg \lla  \min\{\deg(m) \mid (m, (\u, f))\in \G$ has not been lifted$\}$

\While{$\mindeg <= \liftdeg$}
{

$\todo \lla \{\brm \in \G \mid \deg(\brm) = \mindeg$, $\brm$ has not been lifted$\}$

$\H \lla \lift(\todo, \G)$

$\append(\H, \G)$

$\P \lla \eliminate(\H)$

$\update(\P, \G)$

$\liftdeg \lla \max\{\deg(\brm) \mid \brm \in \G$ is primitive$\}$

$\mindeg \lla  \min\{\deg(m) \mid (m, (\u, f))\in \G$ has not been lifted$\}$
}

$\maxcpdeg \lla \max\{\deg(\lcm(\lm(f), \lm(g))) \mid (\lm(f), (\u, f)), (\lm(g), (\v, g)) \in \G\}$

\If{$\maxcpdeg > \liftdeg + 1$}
{
$\liftdeg \lla \maxcpdeg - 1$

\goto 5
}

\return $\{f \mid (m, (\u, f))$ is primitive in $\G\}$

}\caption{The matrix mo-GVW algorithm}
\end{algorithm}

Slightly different from the original mo-GVW algorithm, the matrix mo-GVW algorithm does not include pairs like $(0, (\v, 0))$ in $\G$, because the matrix mo-GVW uses another  technique for checking Syzygy Criterion. This technique is more efficient and will be discussed in Subsection \ref{subsec_checkcriterion}. Note that the matrix mo-GVW only computes a Gr\"obner basis for the ideal $\langle f_1, \ldots, f_m\rangle$.

There are 4 sub-functions in the above algorithm: {\tt lift($\cdot$)}, {\tt append($\cdot$)}, {\tt eliminate($\cdot$)}, and {\tt update($\cdot$)}. Next, we discuss each sub-function one by one.

The function {\tt lift(todo, $\G$)} {\em lifts} each labeled monomial $\brm \in$ {\tt todo} to $x_1\brm$, $x_2\brm$, \ldots, $x_n\brm$, and put all pairs that should be reduced into the result $\H$.

\begin{function}[H]
\DontPrintSemicolon
\SetAlgoSkip{}
\LinesNumbered

\SetKwData{todo}{todo}
\SetKwData{goto}{goto step}
\SetKwFunction{return}{return}
\SetKwInOut{Input}{Input}
\SetKwInOut{Output}{Output}
\SetKwFor{For}{for}{do}{end\ for}
\SetKwIF{If}{ElseIf}{Else}{if}{then}{else\ if}{else}{end\ if}

\Input{\todo, a set of labeled monomials; 

\hspace{1.75cm}$\G$, a set of labeled monomials.}

\Output{$\H$, a subset of $\M$.}

\BlankLine

\Begin{

$\H \lla \emptyset$

\For{each $\brm = (m, (\u, f)) \in \todo$}
{
\For{$i \lla 1, 2, \ldots, n$}
{
\If{$x_i\brm$ collides with $\brm' = (x_im, (\v, g)) \in \G$}
{
$t_f \lla x_im/\lm(f)$

$t_g \lla x_im/\lm(g)$

\If{$t_f\lm(\u) \succ t_g\lm(\v)$ and $x_i\brm$ is not rejected by any of LCM, Syzygy and Rewritten Criterion}
{
$\H \lla \H \cup \{t_f(\u, f)\} \cup \{t_g(\v, g)\}$
} 

\If{$t_f\lm(\u) \prec t_g\lm(\v)$}
{
$\G \lla (\G \setminus \{\brm'\}) \cup \{x_i\brm\}$
}

\If{$t_f\lm(\u) \prec t_g\lm(\v)$ and $\brm'$ is not rejected by any of LCM, Syzygy and Rewritten Criterion}
{
$\H \lla \H \cup \{t_f(\u, f)\} \cup \{t_g(\v, g)\}$
}

} \Else {
$\G \lla \G \cup \{x_i\brm\}$
}
}
}

\return $\H$

}\caption{lift({\tt todo}, $\G$)}
\end{function}
\smallskip

The function {\tt append($\H$, $\G$)} appends $\H$ with the pairs that are used to reduce others.

\begin{procedure}[H]
\DontPrintSemicolon
\SetAlgoSkip{}
\LinesNumbered

\SetKwData{done}{done}
\SetKwInOut{Input}{Input}
\SetKwInOut{Output}{Output}
\SetKwFor{For}{for}{do}{end\ for}
\SetKwIF{If}{ElseIf}{Else}{if}{then}{else\ if}{else}{end\ if}

\Input{$\H$, a subset of $\M$; 

\hspace{1.75cm}$\G$, a set of labeled monomials.}

\BlankLine

\Begin{

$\done \lla  \{\lm(h) \mid (\w, h) \in \H\}$

\While{$\exists\, m \in \{$monomials in $h \mid (\w, h)\in \H\} \setminus \done$}
{
$\done \lla \done \cup \{m\}$

\If{$\exists\, \brm' = (m, (\v, g)) \in \G$}
{
$\H \lla \H \cup \{(m/\lm(g)) (\v, g)\}$
}
}

}\caption{append($\H$, $\G$)}
\end{procedure}

The function {\tt eliminate($\H$)} does reductions to pairs in $\H$ in the following way. 

\begin{function}[H]
\DontPrintSemicolon
\SetAlgoSkip{}
\LinesNumbered

\SetKwData{todo}{todo}
\SetKwData{goto}{goto step}
\SetKwFunction{return}{return}
\SetKwInOut{Input}{Input}
\SetKwInOut{Output}{Output}
\SetKwFor{For}{for}{do}{end\ for}
\SetKwIF{If}{ElseIf}{Else}{if}{then}{else\ if}{else}{end\ if}

\Input{$\H$, a subset of $\M$.}

\Output{$\P$, a subset of $\M$.}

\BlankLine

\Begin{

$\P \lla \emptyset$

\For{each $(\u, f) \in \H$}
{
$\H \lla \H \setminus \{(\u, f)\}$

\If{$\exists\, (\v, g) \in \H \cup \P$ s.t. $\lm(f) = \lm(g)$}
{
\If{$\lm(\u) \succ \lm(\v)$}
{
$\H \lla \H \cup \{(\lc(g)\u-\lc(f)\v, \lc(g)f - \lc(f)g)\}$
}
\If{$\lm(\u) \prec \lm(\v)$}
{
$\H \lla (\H \setminus \{(\v, g)\})  \cup \{(\lc(f)\v - \lc(g)\u, \lc(f)g - \lc(g)f)\} \cup \{(\u, f)\}$
}
}
\Else
{
$\P \lla \P \cup  \{(\u, f)\}$
}

}

\return $\P$

}\caption{eliminate($\H$)}
\end{function}

The function {\tt eliminate($\H$)} is generally done by using linear algebra. Specifically, first, we sort pairs in $\H$ with an ascending ordering on signatures. Second, polynomials are converted to rows of a matrix. Third, we compute the echelon form of this matrix by using one-side elimination such that rows with higher signatures can only be reduced by rows with lower signatures. At last, we convert rows of the matrix to polynomials.

The function {\tt update($\P, \G$)} collects new labeled monomials and appends them to $\G$.

\begin{procedure}[H]
\DontPrintSemicolon
\SetAlgoSkip{}
\LinesNumbered

\SetKwInOut{Input}{Input}
\SetKwInOut{Output}{Output}
\SetKwFor{For}{for}{do}{end\ for}
\SetKwIF{If}{ElseIf}{Else}{if}{then}{else\ if}{else}{end\ if}

\Input{$\P$, a subset of $\M$; 

\hspace{1.75cm}$\G$, a set of labeled monomials.}

\BlankLine

\Begin{

\For{each $(\w, h) \in \P$ with $h \not= 0$}
{
\If{$\exists\, \brm' = (m', (\v, g)) \in \G$ s.t. $m' = \lm(h)$}{

\If{$\lm((m'/\lm(g))\v) \succ \lm(\w)$}
{
$\G \lla (\G \setminus \{\brm'\}) \cup \{(\lm(h), (\w, h))\}$
}

} \Else { $\G \lla \G \cup \{(\lm(h), (\w, h))\}$ }

}

}\caption{update($\P$, $\G$)}
\end{procedure}


\subsection{Checking Criteria efficiently} \label{subsec_checkcriterion}

Clearly, LCM criterion can be checked directly in the function {\tt lift}$(\cdot)$.

In Subsection \ref{subsec_mosgb}, we give a general definition of Syzygy Criterion. That is, a labeled monomial $(m, (\u, f))$ is rejected by Syzygy Criterion w.r.t. $\G$, if there exits $(0, (\v, 0)) \in \G$ such that $\lm(\v)$ divides $t\lm(\u)$ where $t = m/\lm(f)$. Generally, finding such $(0, (\v, 0)) \in \G$ needs to traverse many labeled monomials in $\G$, which may cost much time.

To check Syzygy Criterion efficiently in the matrix mo-GVW algorithm, similarly as done in matrix-F5, we only use principal syzygies instead of all syzygies. 

\begin{cor}[{\bf Principal Syzygy Criterion}]
Let $(m, (\u, f))$ be a labeled monomial with $m\not=0$ and $\lm(\u) = x^\alpha\e_i$. Let $t_f = m/\lm(f)$ and $\G$ be a subset of labeled monomials. If there exists $(t_f x^\alpha, (\v, g)) \in \G$ with $\lm(\v) = x^\beta\e_j$ such that $\e_i \succ \e_j$, then the labeled monomial $(m, (\u, f))$ does not need to be mutural-reduced.
\end{cor}

If such $(t_f x^\alpha, (\v, g))$ exists in $\G$, then $(0, (g\e_i - f_i\v , 0))$ is a syzygy labeled monomial since $(g\e_i - f_i\v)\cdot \f = g f_i - f_i g = 0$.  Besides, we have assumed that $\prec_s$ is a position over term extension of $\prec_p$, so $\e_i \succ \e_j$ implies that $\lm(g\e_i - f_i\v) = \lm(g)\e_i$ divides $t_f x^\alpha\e_i = t_f \lm(\u)$. Therefore, if $(m, (\u, f))$ is rejected by Principal Syzygy Criterion, it can also be rejected by Syzygy Criterion if $(0, (g\e_i - f_i\v , 0))\in \G$.

Using Principal Syzygy Criterion instead of Syzygy Criterion may lead to some undetected redundant computations, but checking Principal Syzygy Criterion  is much more efficient then checking Syzygy Criterion, particularly in complicated systems. So in the implementation of mo-GVW, we prefer to using Principal Syzygy Criterion.

Regarding to Rewritten Criterion, we can check Rewritten Criterion during the function {\tt lift}$(\cdot)$, but this may be not so efficient sometimes. An alternative way is to check Rewritten Criterion during the sort of pairs in $\H$ after the function {\tt lift}$(\cdot)$ is over. Because when we are sorting pairs in $\H$ by their signatures, it is easy for us to find two pairs having the same signature. In this case, we can discard one of them directly based on Rewritten Criterion. 

\subsection{Other details} \label{subsec_others}


There are many cases that we need to check whether an object belongs to a large set of objects, including
\begin{itemize}

\item Line 5 of the function {\tt lift($\cdot$)}, 

\item Line 3 and 5 of the procedure {\tt append($\cdot$)}, 

\item Line 3 of the procedure {\tt update($\cdot$)}, and 

\item the implementation of Principal Syzygy Criterion.

\end{itemize}
For these cases, instead of traversing objects in the large set, we can find out whether the desired object lies in the large set  by using a hash table. Similar method is used in \citep{SunDW13}.

A flag should be designated to each labeled monomial in $\G$ in order to show whether this labeled monomial has been lifted. We also need flags to avoid inserting duplicated pairs into $\H$ in the function {\tt lift($\cdot$)}.


\section{Experimental results} \label{sec_timming}

We implemented the mo-GVW algorithm over boolean polynomial rings in C++. The elimination of matrices is mainly done by linear algebraic routines for dense matrices over $GF(2)$. These routines include the function $gvw\_ple()$ and some other efficient routines from the library M4RI \citep{Albrecht13}, where $gvw\_ple()$ is used for eliminating matrices in signature-based algorithm and it is reported in \cite{SunDW14}. 

We tested several square boolean polynomial systems generated by Courtois in \citep{Courtois13} and a few HFE systems downloaded from \citep{Steel04}. The system $n\times n$ means that the input square polynomial system has $n$ polynomials in $n$ variables. When computing Gr\"obner bases for HFE systems, techniques of dealing with mutant pairs in \citep{SunDW14} are used. The ordering $\prec_p$ is the Graded Reverse Lexicographic ordering, and $\prec_s$ is a position over extension of $\prec_p$. The experimental platform is MacBook Pro with 2.6 GHz Intel Core i7, 16 GB memory.

We tested our implementation of mo-GVW with our previously implemented M-GVW \citep{SunDW14} and some intrinsic Gr\"obner basis functions on public softwares for solving the above systems. 
The computing times in seconds are listed in Table 1 and 2.

{
\begin{table}[H]\centering
\medskip
\begin{tabular}{c c c c c c}
\hline
Syst. & Maple & Singular & Magma & M-GVW & mo-GVW\\

& (ver. 17) & (ver. 3-1-6) & (ver. 2.20-3) & (\citep{SunDW14}) & \\ \hline

$16 \times 16$ & 4.088 & 5.210 & 0.130 & 0.543 & 0.076 \\

$17 \times 17$ & 9.891 & 12.886 & 0.230 & 0.895 & 0.124 \\

$18 \times 18$ & 22.340 & 31.590 & 0.950 & 1.588 & 0.219 \\

$19 \times 19$ & 48.314 & 84.771 & 0.860 & 2.728 & 0.374 \\

$20 \times 20$ & 107.064 & 265.325 & 1.000 & 4.664 & 0.646 \\

$21 \times 21$ & 218.479 & 724.886 & 2.670 & 8.226 &1.338 \\

$22 \times 22$ & 839.067 & $>1$h & 7.410 & 28.840 & 4.178 \\

HFE\_25\_96 & 121.681 & $>1$h & 1.160 & 3.418 & 0.881 \\

HFE\_30\_96 & 619.745 & $>1$h & 2.550 & 15.168 & 3.634 \\

HFE\_35\_96 & 2229.239 & $>1$h & 6.950 & 57.988 & 11.688 \\

\hline
\end{tabular}\\
\caption{\small  Maple, Singular, Magma, and M-GVW vs mo-GVW}
\end{table}
}

{
\begin{table}[H]\centering
\medskip
\begin{tabular}{c c c c c c c}
\hline
Exam. & $23 \times 23$ & $24 \times 24$ & $25 \times 25$ & $26 \times 26$ & $27 \times 27$ & $28 \times 28$ \\ \hline

Magma(ver. 2.20-3) & 15.630 & 100.600 & 139.100 & 306.570 & 560.150 & 1169.150 \\

mo-GVW & 7.622 & 102.032 & 226.752 & 472.561 & 946.823 & 1882.418 \\

\hline
\end{tabular}\\
\caption{\small  Magma vs mo-GVW}
\end{table}
}

From the above tables, we can see that our implementation of mo-GVW outperforms the implementation of M-GVW \citep{SunDW14}, and it is also very efficient for systems that have relative small size, but is not so efficient as Magma for relative large systems and HFE systems. We think this is because the following reasons. First, the M-GVW algorithm uses a similar structure to the algorithm presented in \citep{Albrecht10}. Mo-GVW uses a frame like XL and avoids generating J-pairs, and mo-GVW also uses an improved method of constructing matrices \citep{SunDW13}, so mo-GVW outperforms M-GVW. Second, the elimination of matrices in present implementation of mo-GVW is mainly  done by linear algebraic routines for dense matrices. When the size of systems becomes larger, the matrices generated during the computations become sparser, so our implementation of mo-GVW becomes less efficient than Magma. Third, in the tested HFE systems, linear polynomials always appear after eliminating three 4-degree matrices (i.e. matrices corresponding to 4-degree polynomials). The eliminating results in the first 4-degree matrix should be used to speed up the elimination of the second and third 4-degree matrices. However, this is a bit difficult to be done in our present implementation of  mo-GVW, since only dense linear algebraic techniques are used now.

\section{Conclusions} \label{sec_conclusion}

A new frame of the GVW algorithm is presented in this paper. The new algorithm is called a monomial-oriented GVW algorithm or mo-GVW algorithm for short. Being different from the original GVW algorithm, mo-GVW makes efforts to find the smallest signatures for given monomials. By using this new frame, mo-GVW avoids generating J-pairs, and also provides efficient manners to find reducers and check criteria. We implemented the mo-GVW algorithm over boolean polynomial rings. The experimental results show that mo-GVW is very efficient when the systems are not very complicated.

However, many aspects in the implementation of mo-GVW can still be improved further. The most important one is that the implementation should be improved by using sparse linear algebraic techniques, because matrices generated during the Gr\"obner basis computations are very sparse. This will be our main work in the future.

\end{document}